\documentclass[a4paper]{article}
\usepackage{listings}
\usepackage{xcolor}
\usepackage{multirow}
\usepackage{jheppub}  
\usepackage{natbib}
\usepackage{dcolumn}
\usepackage[english]{babel}
\usepackage[utf8x]{inputenc}
\usepackage[T1]{fontenc}
\usepackage{palatino}
\usepackage{amsmath}
\usepackage{afterpage}
\usepackage{graphicx, subcaption}
\usepackage[colorinlistoftodos]{todonotes}
\usepackage[colorlinks=true]{hyperref}
\usepackage{makeidx}
\newcommand{\be}{\begin{equation}}  
\newcommand{\ee}{\end{equation}}  
\newcommand{\bea}{\begin{eqnarray}}  
\newcommand{\eea}{\end{eqnarray}}  
\makeindex

\begin{document}

\vspace*{1.2cm}

\thispagestyle{empty}
\begin{center}

{\LARGE \bf A Forward Multiparticle Spectrometer for the LHC: Hadron spectra and Long-lived particle search}

\par\vspace*{7mm}\par

{

\bigskip

\large \bf Michael G. Albrow }

\bigskip

{\large \bf  E-Mail: albrow@fnal.gov}

\bigskip

{Fermi National Accelerator Laboratory, Batavia, IL 60510, USA}

\bigskip

{\it Presented at the Workshop of QCD and Forward Physics at the EIC, the LHC, and Cosmic Ray Physics in Guanajuato, Mexico, November 18-21 2019}

\vspace*{15mm}

{  \bf  Abstract }

\end{center}
\vspace*{1mm}

\begin{abstract}

I describe a possible Forward Multiparticle Spectrometer (FMS) that could be installed downstream of the 
superconducting recombination dipole D1 in Run 4, between $z$ = 96 m - 126 m to measure multi-TeV hadron spectra in low luminosity 
pp collisions
at $\sqrt{s}$ = 14 TeV, as well as p+O and O+O collisions as relevant for cosmic ray showers. Light antinuclei and charmed 
hadrons at high Feynman $x_F$ can be measured, both of importance for astrophysics. At the full high luminosity HL-LHC a 
search for new long-lived neutral particles (LLPs) decaying in a 20 m long, 70 cm diameter vacuum pipe to visible decay modes
(including $\gamma\gamma, e^+e^-, \mu^+\mu^-, \tau^+\tau^-, c\bar{c}$ and jets) can be made. The FMS is especially well suited for LLPs 
with 1 GeV $< M(X)<$ 10 GeV and lifetimes $c \tau $ from about 10 m to several km.

 I discuss this as a possible
addition to CMS but it has no formal approval yet, therefore the talk is not given ``on behalf of CMS''.
\end{abstract}
  
\section{Introduction}


The sparcity of accelerator data on particle production in the forward direction above $\sqrt{s}$ = 63 GeV at the 
CERN Intersecting Storage Rings (ISR), its importance for understanding cosmic ray showers, and the possibility of
measurements at the LHC was addressed in Ref. \cite{albrowcr}.

   We have been developing a forward multiparticle spectrometer, FMS, that could be added as a new subsystem to CMS for 
Run 4 (2027+). A schematic overview of the spectrometer is shown in Figure 1.
The main detectors are situated at $z$ = 116 m - 126 m and surround the beam pipe between radii 
$R_{in}$ = 12 cm and $R_{out}$ = 35 cm\footnote{All dimensions are provisional and subject to optimisation.}.
The  \textsc{left+right} and \textsc{up+down} azimuthal regions have distinct physics motivations and operational modes;
 hadron spectroscopy and a new long-lived particle (LLP) search respectively. 
The detectors can use the same techniques as the CMS Endcap upgrade planned for Run 4, with silicon tracking 
and calorimetry with precision timing, followed by a magnetised toroid with GEM layers for muon measurement. 
Transition radiation detectors (TRD) for TeV hadron identification are being developed \cite{romaniouk}; 
these are the only detectors not included in
the CMS upgrade plans. They are essential for the hadron mode, but optional for the LLP mode.
The area is only about 0.3 m$^2$, which is less that 1\% of the future Endcap. 

An earlier talk on the hadron mode is given in Ref.\cite{vfhsmga}.
Forward spectra of $\pi^\pm, K^\pm, p$ and $\bar{p}$ have not been measured above
$\sqrt{s}$ = 63 GeV at the ISR \cite{sasisr1,sasisr2} but are important to understand cosmic ray showers. 
At the LHC with $\sqrt{s}$ = 14 TeV we will be 220 times higher in $\sqrt{s}$. In fixed target
 terms, as appropriate for cosmic ray showers, $E_{BEAM}$ is about 50,000 times higher. 
The ISR energy is well below the famous knee in the cosmic ray spectrum; the LHC energy is 
well above. An excess of muons is observed in very high energy showers compared with expectations; forward spectra
at the LHC may shed light on this, as well as being relevant for atmospheric neutrinos, which are a background to
cosmic neutrinos as seen in ICECUBE. 

Event generators such as \textsc{pythia} have not been tuned for this region
 since there is little data. Low $p_T$ physics being non-perturbative QCD is theoretically more challenging than high $p_T$
and is worthy of more attention; this is the ''low-$Q^2$ frontier'' of QCD.
At the LHC only leading protons with Feynman-$x_F \gtrsim$ 0.9 and neutral particles
 (mainly $\pi^0$ and neutrons) at $\theta = 0^\circ$ have been measured \cite{zerodeg}, demonstrating
 the very large spread in cosmic ray shower Monte Carlos. 

When planning future high energy hadron colliders, such as the $\sqrt{s}$ = 100 TeV 
FCC (p+p mode, as well as with ions) predictions for 
radiation levels in the forward direction
should benefit from improved knowledge of these cross sections.
\section{Description of D1 (81m) to TAXN (127m) region}

The straight section downstream of IR1 (ATLAS) and IR5 (CMS) between the end of the new superconducting D1 dipole at $z$ = 81 m 
and the entrance to the TAXN absorber at 126.5 m is mostly free of equipment, with a new straight beam pipe presently planned to 
have $R$ = 7.5 cm at the front increasing to 12.5 cm at the back. (The regions downstream of LHCb and ALICE are where 
the proton beams are injected and are more complicated.) We propose to change the design of this pipe to have an 
enlarged radius, nominally $R$ = 40 cm over at least 20 m, from $z$ = 96 m
 to 116~m. Immediately downstream of D1 there is to be a cold diode structure (in DFBX) parallel to the beam pipe 
which, if it cannot be repositioned, limits the beginning of the proposed detector system, shown schematically in Fig. 1.

The first new element is an iron (ASII 1010 low-carbon steel) toroid (I thank V. Khashikhin, Fermilab, for the study), a cylinder of length 3 m, 
$R_{in}$ = 8 cm and $R_{out}$ = 40 cm. It is constructed in two halves for easy assembly/disassembly and allowing 
separation of top and bottom halves for bakeout of the beam pipe.
Two water-cooled copper coils, both in the bottom half, with currents of 5 kA each, 
give a circular field in the iron varying from 1.9 T at the inner radius to 1.5 T at the outer. 
All charged particles (mostly muons) exiting the toroid steel are measured in a counter hodoscope mounted on the back of the steel,
 followed by track chambers, e.g. a pair of GEMs or silicon strip layers, separated by 1 m. 
The field deflects charged particles emerging from the 
back of D1 inwards or outwards, reducing the flux of muons at the detectors downstream. This is predicted by 
\textsc{fluka} to be 0.9 (0.65)
per bunch crossing even with 140 interactions (HL) without (with) the toroid powered.

The field at the center of the beam pipe is less than 3 Gauss; both incoming and outgoing beams are inside the pipe but not centered.
 If necessary, the field inside the pipe could be 
reduced by a thin iron shield around it. In the charged 
hadron spectroscopy mode, the 
main role of this toroid is additional background reduction; the particles to be measured pass through the central hole.

 \begin{figure}[t]
  \vspace{-0.5 in}
 \begin{center}
\makebox[\textwidth][c]{\includegraphics[angle=270,origin=c,width=200mm]{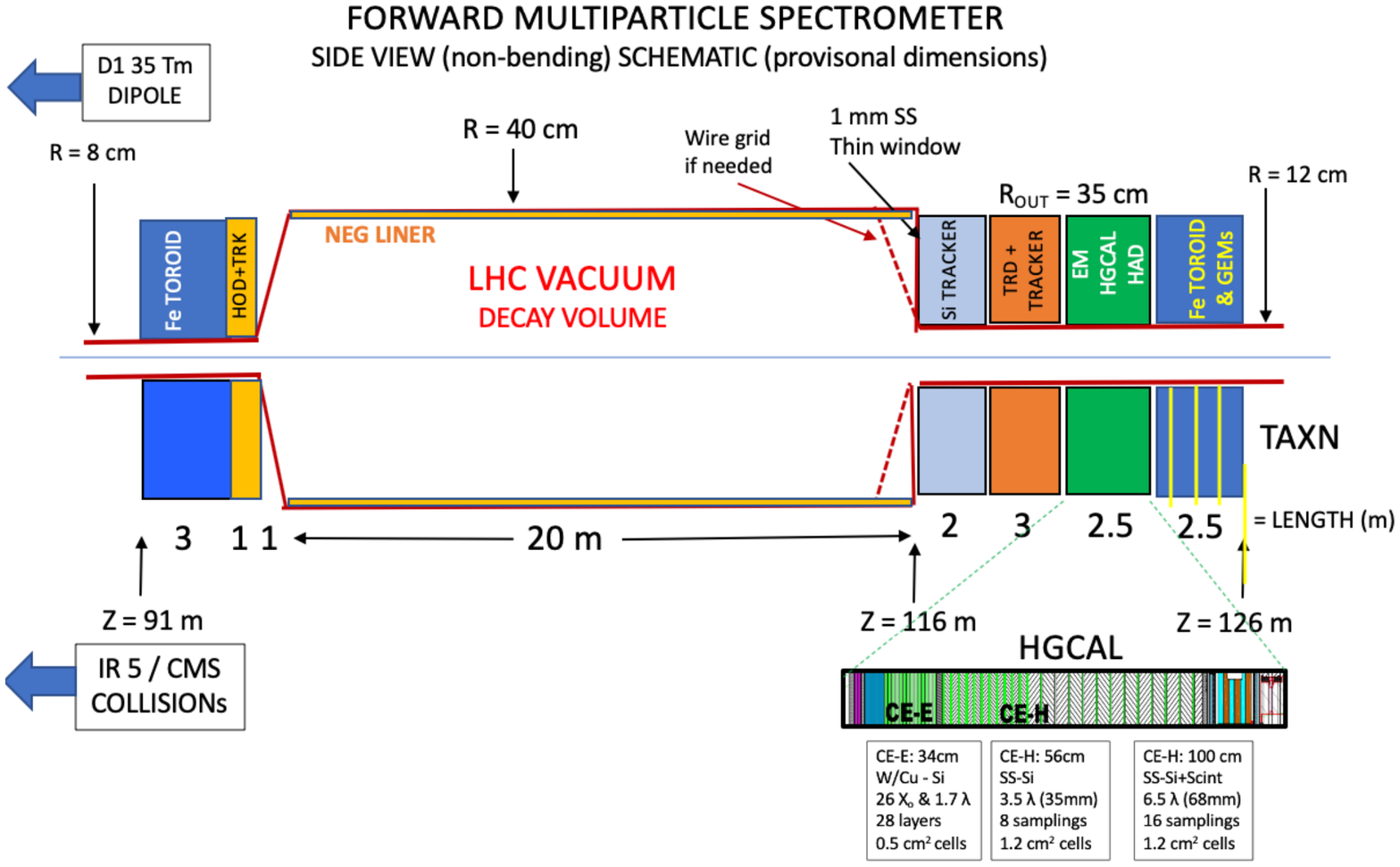}}
\end{center}
 \vspace{-2.2in}

\caption{Schematic layout of proposed FMS spectrometer (side view). Dimensions are subject to optimization, in particular the 
length budget. The start in $z$ could be earlier if the LHC cold diode can be displaced, and the space allocation for the main detectors 
can be increased at the expense of vacuum volume.}
 \end{figure}

Immediately after the toroid+tracker the pipe transitions to a wide pipe, similar to the 25.7 m long pipe  at ALICE in 
LSS2. NEG-coated liners inside an 80 cm pipe leave a clear aperture of diameter 70 cm. (I thank V. Baglin, CERN,
for discussions on the beam pipe.) The main difference 
from the ALICE pipe is that the transition at $z$ = 116 m to the small pipe should be such as to minimize interactions 
and especially multiple scattering. A 1 mm thick steel window perpendicular to the pipe axis gives a multiple scattering 
angle $\theta_\circ = 3 \times 10^{-5}$ for 100 GeV/c particles, decreasing like 1/$p$. An option is to have a thinner window 
with strengthening ribs. To avoid the beam ``seeing'' a sharp change in pipe diameter an internal inclined wire grid or similar 
can be employed\footnote{A detailed design will require an Engineering Change Request from CMS, which has not yet formally considered 
this project.}.

After exiting the steel window the main elements of the spectrometer could use identical technology to the CMS Endcap upgrade, 
namely silicon pixel tracking, followed by electromagnetic and hadron calorimetry based on silicon pads with tungsten/copper 
and steel plates. Precision timing $\sim$ 25 ps is planned, and is important.

The need for precise tracking is very different for the hadron mode and the LLP mode. In the former case (L and R 
quadrants, hadrons of 1 - 3 TeV) we need to measure the momenta of particles coming directly from the collision region 
(or from charm decays) using the 
transport matrix through the magnet lattice, and also to reject beam halo and tracks coming from interactions in the upstream pipe and 
other material.
In the LLP mode the detected particles, from
a decay in the vacuum, have not traversed any magnetic field; the essential need is to project the tracks back to a 
vertex and ensure it is inside the vacuum. Also we need to ensure that the neutral parent points back to the collision region 
through the steel (making allowance for missing neutrinos in any $\tau^+\tau^-$ events!). 
Particles with momenta as low (\emph{sic}) as 50 GeV/c may be of interest. One may dedicate 
about 3 m of space for tracking, with $\sigma \sim$ 20 $\mu$m resolution giving 
angular resolution $< 10^{-5}$. Transition radiation detectors (needed especially for 
hadron spectroscopy, but less essential for the LLP search) can incorporate tracking, 
so one may consider combining them and dedicating up to perhaps 5 m for both silicon strips or pixels and the TRD. 

Transition radiation detectors, sensitive to $\gamma = E/m$, have usually been used to help distiguish electrons 
from pions at low energies. Anatoli Romaniouk and the ATLAS TRD Group have been developing detectors that could 
distinguish $\pi, K,$ and $p$ in the TeV region \cite{romaniouk}. Cherenkov counters are ineffective as $\beta$ is too
close to 1.0.
X-rays are emitted from transitions between media of 
different dielectric constants, or plasma frequencies, with a small probability which rises with $\gamma$ and then saturates.
One can select radiator materials and thicknesses and gap widths to optimise for a selected range. Tests have been done at the 
SPS using layers of xenon-filled straw tubes between different foils with electrons, muons and pions. The yields, X-ray 
energy spectra and angular distributions are very well predicted by detailed simulations. Interestingly the typical emission
angle decreases like 1/$\gamma$, and high granularity silicon or GaAs pixel detectors measuring the emission angles of even a 
few  X-ray photons may improve the separation power. 

The imaging calorimeter (HGCAL) layers will be sensitive to muon tracks, and fast timing will be incorporated (perhaps with 
LGADs) to help with 
background reduction. The EM part of the calorimeter has very good measurement of high energy shower directions, 
addressing the challenge 
of locating the decay point of an $X \rightarrow \gamma\gamma$ decay within the vacuum pipe. This has no physics background; 
probably the main backgrounds are photons from material before the decay
region with a mis-measured vertex, or from two unrelated photons that appear to come from a vertex. 
The tracker + imaging calorimeter combination identifies $\gamma, e, \mu$ and hadrons, and the TRD can provide some 
distinction between $\pi^+\pi^-$ and $K^+K^-$. This capability would be especially powerful in searching for
 $X \rightarrow c \bar{c}$ and $X \rightarrow \tau^+\tau^-$, even $X \rightarrow$ jet + jet. The FMS is particularly well 
suited to LLPs in the M(X) 
= 1 GeV to 10 GeV range (fixed target experiments have higher luminosity and are more sensitive to lighter particles, e.g. 
dark photons with M(A') < 500 MeV).

Behind the calorimeter we propose another iron toroid, identical to the plug at the 
front end except that it is subdivided longitudinally with a few gaps allowing 
insertion of muon tracking layers (e.g. GEMs). With 1.5 T the bending angle for a 100 GeV/c 
muon is 13.5 mrad, compared with the multiple scattering $\theta_{rms}$ = 2 mrad. 
The sensitivity to $X  \rightarrow \mu^+\mu^-$ needs a full simulation, but the 
mass resolution is probably a minor issue, since there is no physics background 
from $K^0$ decays (B.R < $10^{-8}$) or any other SM particles. The back of this toroid is shielded from background 
coming from behind by the TAXN absorber.

Like the front toroid, the back detectors cover full azimuth but can be separated into top and bottom halves, or quadrants.

Installation of an FMS in both outgoing beams is technically possible and would give twice the data and LLP sensitivity for less 
than twice the cost.

\section{Hadron spectroscopy}

When the ISR came into operation in 1971 Feynman proposed that forward hadron
 spectra should scale with energy $\sqrt{s}$
when plotted as a Lorentz-invariant cross section at fixed $p_T$ vs. $x_F = p_z/p_{beam}$; this is Feynman scaling.
It was based on the parton model, pre-QCD, and while it is a good approximation in the ISR energy range for light 
particles at low-$p_T$ \cite{sasisr1, sasisr2},    QCD has
scaling violations, heavy flavors have thresholds, etc. Feynman scaling should not hold over the large
energy range from $\sqrt{s}$ = 63 GeV to 14,000 GeV! 

The new 35 Tm beam recombination dipole D1, ending at $z$ = 81 m, is here used as a
 spectrometer magnet, deflecting charged particles into right (R) and left (L) quadrants. 
The beam crossing angle (planned to be vertical for CMS, 250 $\mu$rad half-crossing angle) and the quadrupoles affect the 
distributions, as shown in Fig. 2. A large
 beam pipe, $R$ = 40 cm, from $z$ = 96 m to 116 m at the end of which is a steel vacuum
 window about 1 mm thick, allows charged particles to enter the spectrometer, where they 
can be measured in short low luminosity p+p, p+O, and O+O runs. The acceptance for primary
 charged particles is approximately $p_z$ = 1 - 3 TeV/c. Higher $p_z$ particles remain within the pipe.
Fig. 2 shows the spatial distribution of primary charged particles at $z$ = 116 m (M. Sabate-Gilarte and F. Cerutti).

Hadrons from fragmentation of diffractively  excited protons, $p \rightarrow p^*$ populate this region,
and in low pileup data it would be interesting to study in combination with a leading proton in the opposite 
direction if there are suitable Roman pots.
 
The FMS can also measure light nuclei and 
 antinuclei ($\bar{d}, \bar{t}, ^3\!\bar{He}$) which are relevant for understanding 
$\gamma$-rays from the galactic center and a possible dark matter annihilation signal. 
It will have acceptance for $J/\psi \rightarrow \mu^+\mu^-$ and 
charmed hadrons, specifically $D^0 \rightarrow K^- \pi^+$, 
$\bar{D^0} \rightarrow K^+ \pi^-$ and $\Lambda_c^+ \rightarrow 
p K^- \pi^+$ at $x_F >\sim$ 0.8; the decay products
 have low enough momenta to be accepted, shown in Figure 3.
Charm production is important for
 understanding ultra-high energy neutrinos and cosmic rays as well as QCD; intrinsic charm, namely
 $c\bar{c}$ in the proton wavefunction, can give a large cross section \cite{brodsky}. 
The challenge of seeing these narrow charm signals on a large combinatorial background drives
 the need for excellent tracking (traversing only vacuum from the collision point to the window at 116 m) and good
 $\pi/K/p$ separation.
Prompt muons can also be measured, subtracting the spectra from $\pi^\pm$ and $K^\pm$
 decays (which will be known) as another measure of $c$- and $b$-production. Note that the
 mean decay length for a 2.5 TeV charged pion(kaon) is 139(18.5) km, and for a 5 TeV $D^0$ it is 33 cm!

 \begin{figure}[t]
  \vspace{-1.3in}
 \begin{center}
\makebox[\textwidth][c]{\includegraphics[angle=270,origin=c,width=200mm]{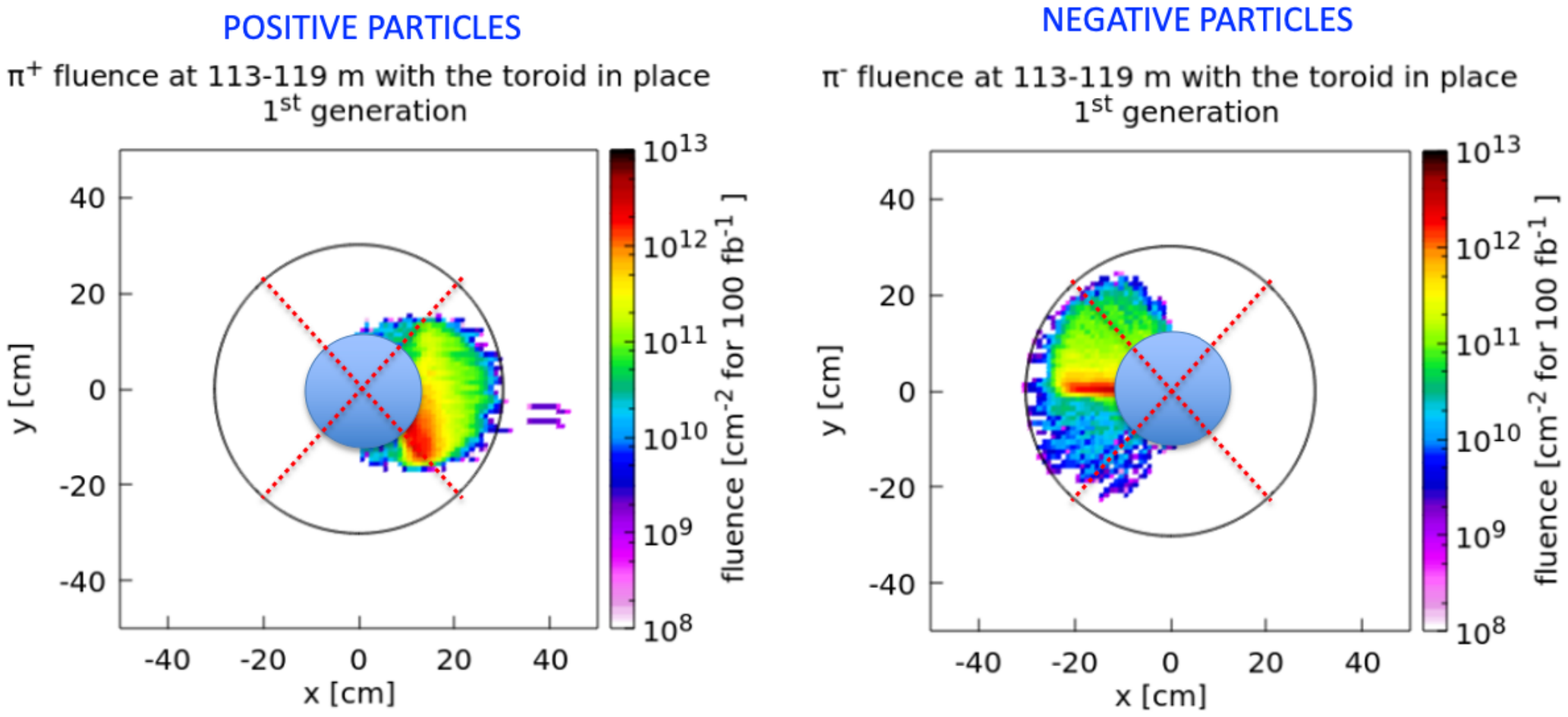}}

\end{center}
 \vspace{-2.7in}

\caption{Fluxes of primary charged pions at $z$ = 116 m per 100 fb$^{-1}$ (M. Sabate-Gilarte).
The central grey disk is the outgoing beam pipe. To convert to numbers per collision divide by 8$\times 10^{15}$.
The regions above and below the pipe are clear of primary particles. The outer radius is now planned to be larger than
shown, namely 35 cm or 40 cm.
}
 \end{figure}

The expected fluxes of charged particles as well as charmed hadrons have been calculated using different 
cosmic ray Monte Carlos by H. Menjo (priv. comm.)
and by M. Sabate-Gilarte (priv. comm. and \cite{april}) using \textsc{fluka} with \textsc{dpmjet} including upstream 
interactions. There is no space here 
for details but e.g. the expected flux of $\mu^\pm$ within $R$ = 30 cm at $z$ = 116 m is only 0.9 per bunch crossing with 
140 interactions, reducing to 0.65 with the front toroid powered, and nearly all of 
these have $p_{\mu}$ < 50 GeV/c. Most pp collisions
produce no direct hadrons in the FMS acceptance (the average is about 0.2) and measurements of the inclusive charged 
hadron spectra could be made with some
pileup, but for multiparticle states like $D^0$ decays the signal:background may be unacceptable unless there is not much pileup.

If there is a zero-degree calorimeter (ZDC/LHCf) between the beam pipes downstream of the FMS to detect neutrons and $\pi^0$, 
we will be able to study coincident events, e.g.  $p \rightarrow n \pi^+ \pi^0$ by diffraction dissociation.

 \begin{figure}[t]
  \vspace{-1.3in}
 \centering
\makebox[\textwidth][c]{\includegraphics[angle=0,origin=c,width=130mm]{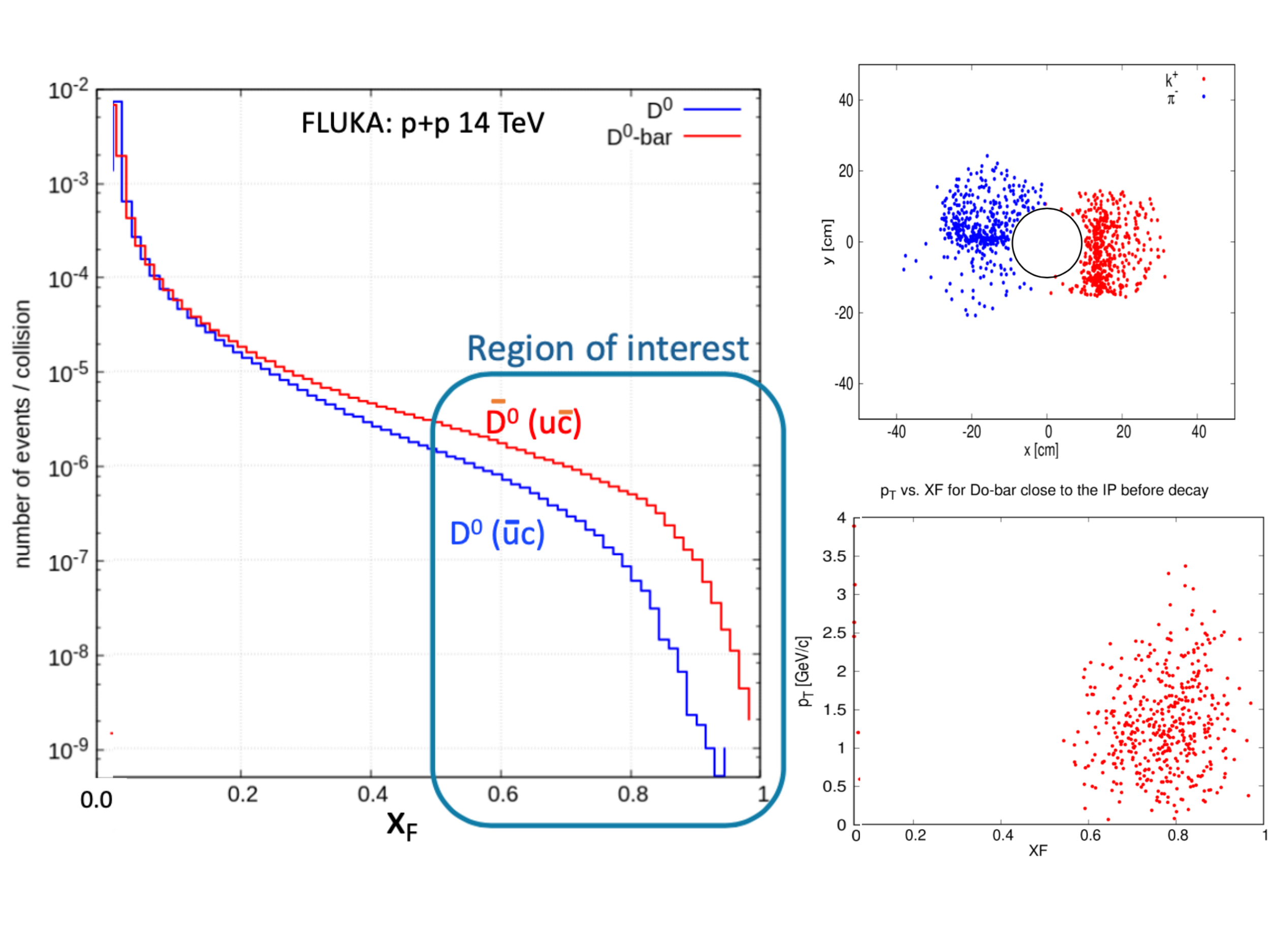}}
 \vspace{-0.4in}

\caption{Left: Spectra in $x_F$ of $D^0$ and $\bar{D^0}$ from FLUKA, p + p at $\sqrt{s} =$ 14 TeV. Top right: Spatial distribution
of $K^+$ and $\pi^-$ from $\bar{D^0}$ decays at $z$ = 116 m. Bottom right: Distribution in the $p_T:x_F$ plane of $\bar{D^0}$
with both $K$ and $\pi$ in FMS acceptance. (M. Sabate-Gilarte)
}
 \end{figure}

\section{Long-lived particle (LLP) search}

In the absence of a discovery of high mass dark matter particles, searches are turning to the possibility that they are 
light (e.g. M(X) $<$ 10 GeV) but weakly interacting. There may be ``portals`` that couple SM particles to dark matter particles, 
that are weak 
enough to penetrate a lot of matter but then decay to known Standard Model particles such as $\gamma\gamma, e^+e^-, \mu^+\mu^-, 
\tau^+\tau^-, c \bar{c}$ and $b\bar{b}$. The FMS can search at full luminosity for 
penetrating neutrals with all of these decays occurring inside the vacuum pipe.
 A \textsc{fluka} calculation by M. Sabate-Gilarte predicts, exiting the front toroid steel, per bunch crossing with 150 
inelastic interactions, 
 0.6 photons, 0.45 neutrons, 0.15 antineutrons, and 0.12 $K^0$ 
above 50 GeV/c. Above 200 GeV/c the fluxes are much less, see Figure 4. 
The upper (U) and lower (D) quadrants are devoid of primary charged particles since 
D1 acts as a sweeping magnet, and the 
detector area is out of the angular range for direct neutral particles. The low occupancy in these quadrants 
 provides an excellent opportunity to search for BSM long-lived neutral particles (LLPs) 
from the primary collisions that penetrate 35 - 50m of steel (> 190 $\lambda_{int}$) 
\footnote{I thank Francesco Cerutti for the following numbers of interaction lengths, for a straight
track from the IP to the end of the D1 cold mass at 81 m: 
320 $\lambda_{int}$ at y = 15 cm, 220 $\lambda_{int}$ at y = x = 15 cm due to yoke holes at 45$^\circ$, 300 $\lambda_{int}$
at y = 20 cm due to the smaller section of the multipole correctors, and 190 $\lambda_{int}$ at y = 10 cm due to the part of the 
path in the vacuum.} 
 in the Q1-Q3 magnets and D1, and decay in the vacuum of the 
large pipe. The decay products, be they photons, electrons, muons or charged hadrons, can be measured in FMS during high 
luminosity running. Excellent tracking to show that the vertex is inside the 20 m-long vacuum region, 
 and not initiated by a charged particle (e.g. $\mu$), should eliminate backgrounds; Standard Model LLPs such as $K^0$ 
and $\Lambda$ are recognized in the spectrometer which has tracking, calorimetry and muon chambers, and can be reduced using
mass and lifetime information.

One may ask about of the sensitivity of the FMS to semi-weakly interacting LLPs that do \emph{not} decay in the 
vacuum pipe or interact in the steel absorber but interact inside the calorimeter.
At first sight it may seem that backgrounds are overwhelming, but the HGCAL will provide 
sensitivity to single charged particles, detailed 
shower starting point and directional information, precision timing and  energy measurement. 
The muon chambers immediately behind the calorimeter have information about possible muon content in the shower.
Combining time-of-flight and 
energy measurement gives M(X); look for a peak!
For example, if M(X) = 2 GeV/c$^2$ and p = 50 GeV/c, the flight 
time to the calorimeter is 200 ps later than that of a neutron; for M(X) = 5(10) GeV/c$^2$ with p = 100 GeV/c it is
 more than 0.4(2.0) ns later. While the time resolution on a shower should be $\sim$ 20 ps, the time spread of the collisions
themselves, projected in the forward direction, may be a limiting factor.
Also for decaying LLPs the time-of-flight from collision to signals in the tracker and HGCAL
can be useful information if $\gamma \lesssim$ 20.
  Selecting showers from interacting neutral 
particles emerging from the back of the 20 $\lambda_{int}$ absorbers, and 
(critically) pointing back to the collision region, one might reduce background from scattered high energy 
neutrons to an acceptable level. This unique capability of FMS merits a detailed study.

\subsection{Comparison with FASER}
There are several other experiments searching for penetrating then decaying long-lived particles. 
Here I only compare with that most
similar to FMS, FASER, which is approved for a first run in Run 3 with a decay volume length only 1.5 m and radius 10 cm, 
but is planned to be upgraded to
5 m and 1 m radius for Run 4. It is much further downstream (of IR 1) at $z$ = 480 m - 485 m, after about 100 m of rock absorber.
It is centered on the collision axis, and with a radius $R$ = 0.1(1.0) m has pseudorapidity $\eta$ above 9.2(6.9) neglecting 
beam crossing angle effects.

 \begin{figure}[t]
  \vspace{-0.2 in}
 \centering
\makebox[\textwidth][c]{\includegraphics[angle=0,origin=c,width=140mm]{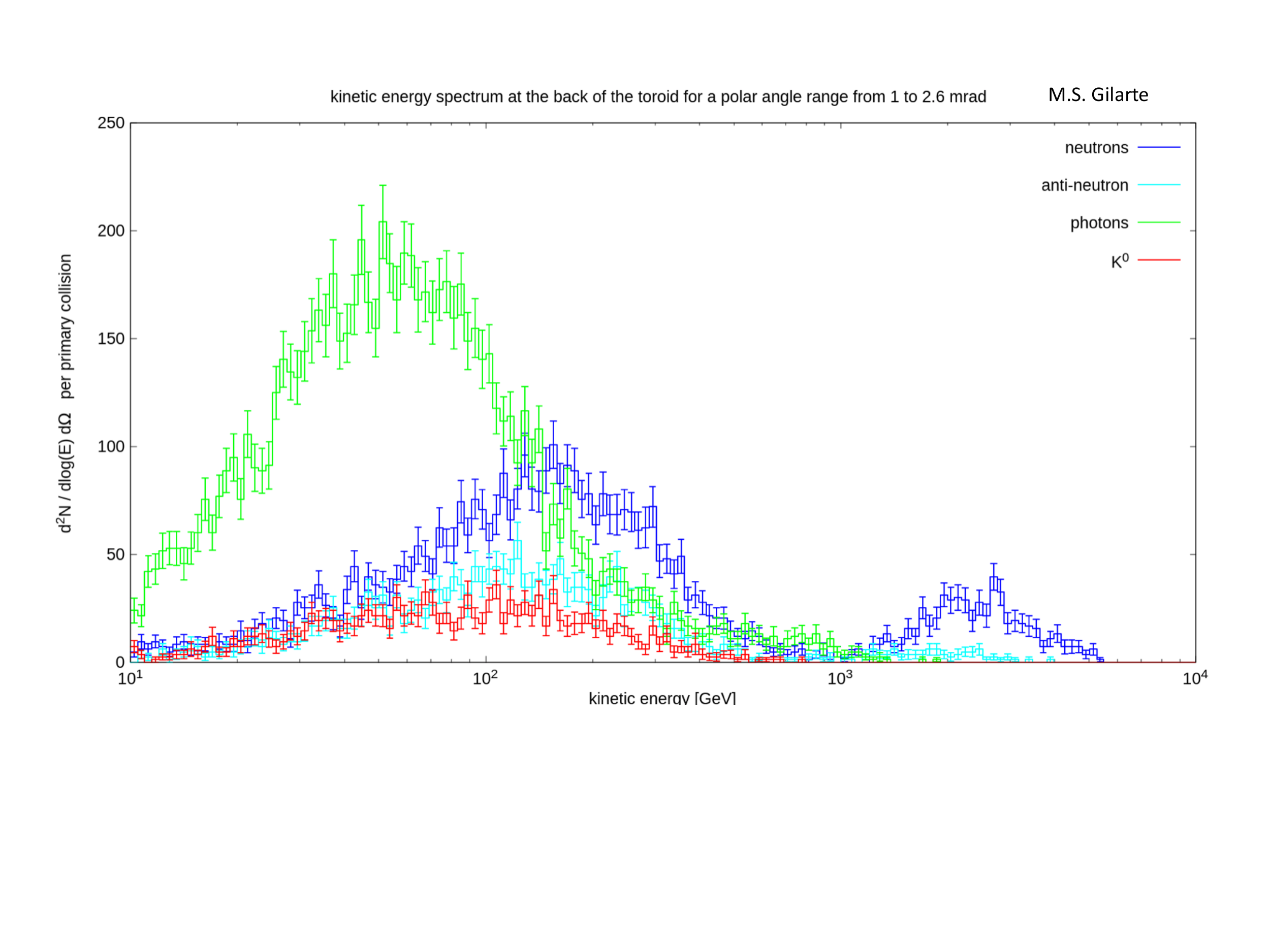}}

 \vspace{-1.0 in}

\caption{Fluxes of neutral particles emerging from the back of the front toroid (M. Sabate-Gilarte).}
 \end{figure}

The probability of a particle that enters a decay volume decaying in it is:
 
\begin{equation}
F = e^{-z_{in}/(\gamma c \tau)} - e^{-z_{out}/(\gamma c \tau)} 
\end{equation}

For FASER(5m) this exceeds $10^{-3}$ for  $\gamma c \tau$ between 130 m and 20 km, and  has a maximum when $\gamma c \tau$ = 480 m 
at which $F = 3.8 \times 10^{-3}$.
The FMS is both closer to the IP and is much longer, and $F$ exceeds 1\% between $\gamma c \tau$ = 24 m and 1.85 km, with a 
maximum of 6.9\% at $\gamma c \tau$ = 116 m.

Since FMS(LLP) with $6.65 < |\eta| < 7.7$ is at a larger polar angle than FASER, we should also 
compare the fluxes of particles as a function
of momentum. These have been calculated by F. Cerutti and M. Sabate-Gilarte with \textsc{fluka} and two examples are shown in Figure 5.
Without having predictions for the production of LLPs of various masses, we assume the spectrum of an A' light enough to come
from $\pi^0$ decays to be similar to that of  $\pi^0$ themselves, and that of an LLP with M(X) $\sim$ 2 GeV/c$^2$ to be similar to that of a $D^0$. So these are only indicative, but
show that the FASER flux is higher for $\pi^0$ with momenta above about 1 TeV, but heavier particles have larger mean $p_T$ 
and the charm flux
is much higher in the FMS(LLP) angular region. For the FASER Run 4 proposal their charm flux will be higher, although this prediction 
has very large uncertainty since there is no data. However FMS in the hadron mode will measure very forward charm, 
largely resolving this issue.

Of course if an A' or similar BSM particle is discovered before Run 4, the FMS should be able to study it in a novel way.

The FASER-$\nu$ extension is to measure neutrino interactions in an emulsion stack at the same location. Since the spectra of 
charged pions, kaons and charmed hadrons at large $x_F$ presently have an order of magnitude uncertainty, the 
neutrino cross sections
cannot be measured without knowing those spectra. FMS (hadron mode) will measure these up to $x_F \sim$ 0.4, thus 
providing a service to the LHC forward neutrino physics program, as well as to experiments studying cosmic neutrinos, such as ICECUBE.

 \begin{figure}[t]
  \vspace{-0.2 in}
 \centering
\includegraphics[angle=0,origin=c,width=150mm]{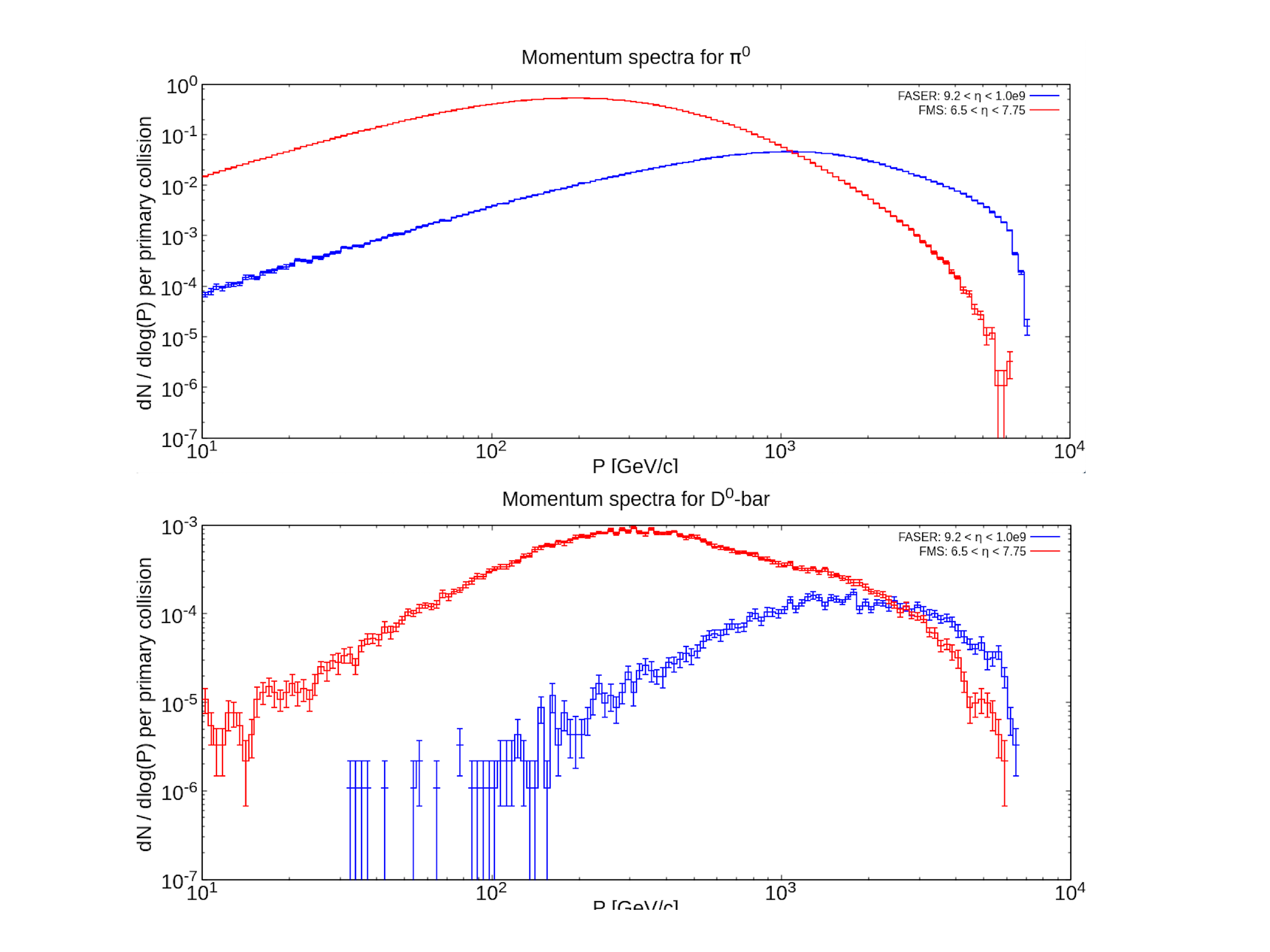}
 \vspace{-0.2 in}

\caption{Spectra calculated using \textsc{fluka} for $\pi^0$ and $D^0$ for FMS ($6.65 < |\eta| < 7.7$)and FASER($\eta > 9.2$ (Run 3).}
 \end{figure}

\section{Hadron interactions}

In the hadron mode, behind the TRD there will be a flux of identified hadrons of known
 momenta between about 1 and 3 TeV/c. One could insert 
thin foil targets, e.g. of carbon and polyethylene (C$_2$H$_4$) with some pixel tracking 
a few meters behind for a special short run.  By counting tracks from a vertex in the foil one could get a 
measure of $\sigma_{inel}$ and the $N_{charged}$ distribution 
for the different beam particles (including light nuclei and antinuclei) on both 
protons and carbon. To make longitudinal space for this the calorimeter  may need to be displaced.

\section{Triggers and data collection}

  In the hadron spectroscopy mode the ideal running condition would be to have an 
average of about one inelastic collision per bunch crossing ($\mu$ = 1), with a 
level-one trigger based on one or more tracks or EM-calorimeter signals. 
The full CMS detector would be read out to study correlations, and the 
single-track rate would need pre-scaling. To maximize statistics for charm etc., 
$\geq 2$-track triggers could include a fast processor selecting candidates. 
While the single-particle inclusive spectra could be measured at higher pileup,
the charm signal:background would become worse; this needs a study. 

In the LLP mode at high luminosity, when a candidate in FMS is accompanied by a 
large number of inelastic collisions, there does not appear to be any value in reading 
out the full central detector with an FMS trigger. The DAQ could then have an FMS-only data stream, with a 
trigger selecting events with charged particles, or an anomalous calorimeter signal, behind the big pipe, 
with no corresponding charged particle  entering at the front.

\section{Conclusions}

A powerful multiparticle spectrometer could be installed in the 30 m straight section 
 between the D1 dipole and the TAXN for physics in Run 4. In the L and R quadrants hadron spectra can be measured in a few days
of low luminosity running. In the U and D quadrants a search can be made at full luminosity for new 
long-lived particles decaying in a large diameter 20 m-long vacuum pipe. The detectors system uses the techniques of the CMS 
Endcap upgrade (but with about 1\% of the area) with novel transition radiation detectors. A longer write-up is in preparation;
new participants are welcome!

\section{Acknowledgements}

I thank the organizers of the Workshop at Guanajuato inviting me and for 
allowing me to include in this write-up the LLP search mode, developed since that workshop. 
I am a CMS member, but this paper is not on behalf of CMS 
but \emph{ad personam} since the project does not yet have any official standing in CMS. I hope that will change. I especially 
want to thank Hiroaki Menjo, Anatoli Romaniouk (TRD development), and CERN staff on LHC: Francesco Cerutti, Marta Sabate-Gilarte and 
Vincent Baglin for information and calculations showing the feasibility of the project.

\end{document}